\documentclass[aps,prl,reprint,amsmath,amssymb,amsfonts,%
footinbib,showpacs,superscriptaddress]{revtex4-1}
\usepackage{graphicx}
\usepackage{color}
\newcommand{\revis}[1]{#1}
\begin{document}
\title{Stabilization of helical \revis{macromolecular} phases by confined
bending}
\author{Matthew J.~Williams}
\email[E-mail: ]{mjw532@uga.edu}
\affiliation{Soft Matter Systems Research Group,
Center for Simulational Physics, The University of Georgia, Athens, GA 30602,
USA}
\author{Michael Bachmann}
\email[E-mail: ]{bachmann@smsyslab.org}
\homepage[\\ Homepage: ]{http://www.smsyslab.org}
\affiliation{Soft Matter Systems Research Group,
Center for Simulational Physics, The University of Georgia, Athens, GA 30602,
USA}
\affiliation{Instituto de F\'{\i}sica,
Universidade Federal de Mato Grosso, Cuiab\'a (MT), Brazil}
\affiliation{Departamento de F\'{\i}sica,
Universidade Federal de Minas Gerais, Belo Horizonte (MG), Brazil}
\date{\today}
\begin{abstract}
By means of extensive replica-exchange simulations of generic coarse-grained
models for helical polymers, we systematically investigate the structural
transitions into all possible helical phases
for flexible and semiflexible elastic polymers with
self-interaction under the influence of torsion barriers. The competing
interactions lead to a variety of
conformational phases including disordered helical
arrangements, single helices, and ordered, tertiary helix bundles.
Most remarkably, we find that a bending restraint entails a clear
separation and stabilization of the helical phases. This aids in
understanding why
semiflexible polymers such as double-stranded DNA tend to form pronounced
helical structures and proteins often exhibit an abundance of helical
structures, such as helix bundles, within their tertiary structure. 
\end{abstract}
\maketitle
Helical segments are ubiquitous secondary structures occurring in
most
\revis{macromolecular systems}. The formation of helical structures is typically attributed to the formation of hydrogen bonds along the backbone of linear polymers, but
it is also known that helices are among the few generic geometries that
a linelike topology can form if an ordering principle (such as a many-body
constraint) is present~\cite{maritan1,vnbj1,vnbj2}. 

\revis{In seminal works, Zimm and Bragg (ZB)~\cite{zimm1,zimm2} showed that the crossover between disordered random coil structures and ordered helical conformations can be described by a  one-dimensional Ising-like model. Therefore, while short-range cooperativity can lead to
structural ordering, in the ZB model this process is not a phase transition
in the strict thermodynamic sense~\cite{scheraga1,badasyan1}. However,
since biologically relevant macromolecules are 
finite systems (on an effectively mesoscopic scale), the
thermodynamic interpretation of structural transitions in such
systems must address finiteness effects accordingly~\cite{mb1}.}

Primary effects of cooperativity can be addressed by generic effective-potential models that allow for the qualitative description of helix--coil transitions~\cite{chen1,rapaport1,carri1,sabeur1}. 
Dominant nonbonded interactions support the formation of tertiary structures including single helices, helix bundles, collapsed globules, or random coils~\cite{cohen1,bocz1,kim1,poulsen1,bbd1,bdb1,bleha1,buehler1}. It has
been shown recently that the alignment of secondary structures
in a tertiary protein fold can be understood as a simple two-state
process~\cite{dill1}.

For a generic flexible polymer chain, in a
crystallization process succeeding the chain collapse, ordered structures 
emerge that are substantially different from tertiary structures known from
realistic biomolecules and typically do not possess secondary
structures~\cite{svbj1,sbj1}. If bending restraints and nonbonded interaction
compete with each other, as it is the case in self-interacting semiflexible
polymers, the ordered structures are known to be rodlike bundles or
toroids~\cite{sslb1}. 

However, less is known about the influence of effective bending restraints
upon transition pathways toward helical structures. A systematic analysis
of the formation and separation of helical phases in phase space in the
presence or absence of bending restraints has not yet been performed. In
this
study, we investigate the relevance of this restraint for the separation
of structural phases by means of replica-exchange Monte Carlo
computer simulations for coarse-grained flexible and semiflexible polymer
models. By scanning the spaces of torsion parameter strength and temperature,
we construct the hyperphase diagrams for entire classes of helical
macromolecules, which allows us to distinguish the different pathways to the
helical folds and enables us to judge the significance of bending
restraints in biomacromolecules.
\begin{figure}
\centerline{
\includegraphics[width=8.8cm]{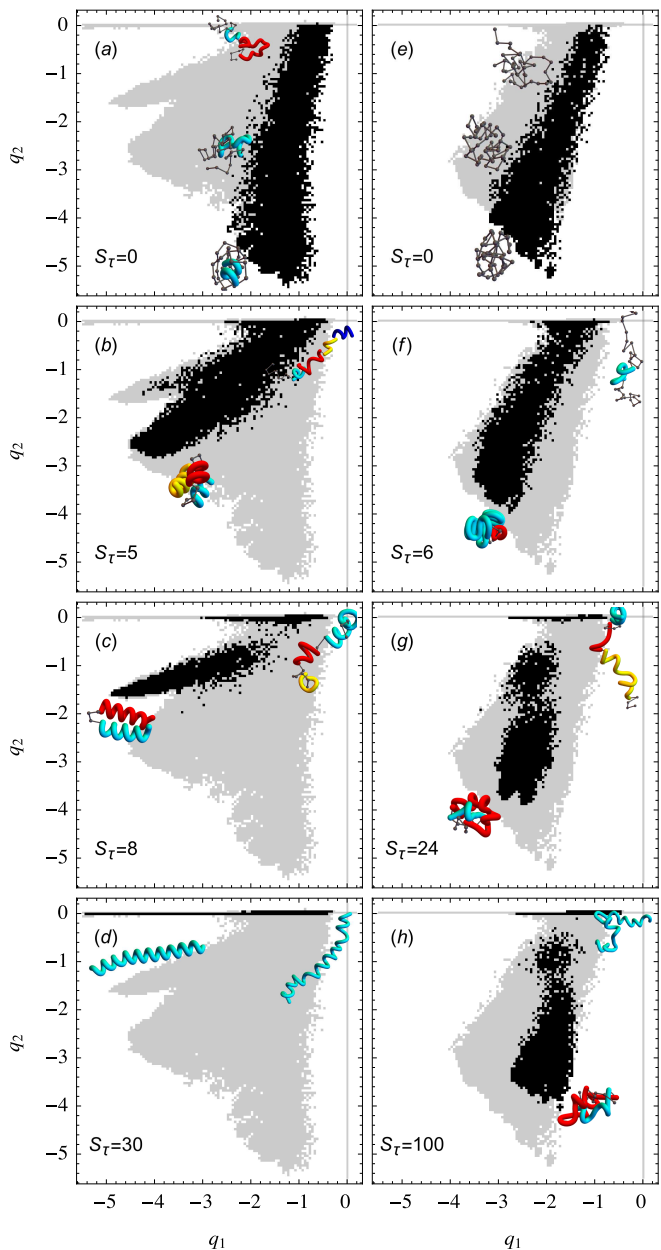}
}
\caption{\label{fig:dist}%
\revis{Scatter plots of conformations of the (a)-(d)
semiflexible (bending restrained) and (e)-(h) flexible (bending
unrestrained) polymers with $40$ monomers in $(q_1,q_2)$ space. 
Lightgray regions represent the generalized ensemble of all conformations
found at all temperatures $T$
and torsion strengths $S_\tau$ simulated. Black regions correspond
to the most populated states at given $S_\tau$ values.} Representative
conformations are shown.}
\vspace*{-5mm}
\end{figure}

We employ a generic coarse-grained bead-spring model for elastic,
self-interacting polymers with torsional interaction. The polymer is
represented by a linear chain of $N$ monomers. 
The
bending energy of flexible
polymers is zero. For
semiflexible polymers, excitations of the bond angle formed by successive
bonds are subject to an energetic penalty. Torsion is induced by an
out-of-plane torsion angle
between three successive bonds. Nonbonded monomers interact
via long-range attractive and
short-range repulsive van der Waals forces, modeled by the Lennard-Jones (LJ)
potential. The energy of a conformation
$\mathbf{X}=\{\mathbf{x}_1,\mathbf{x}_2,\ldots,\mathbf{x_N}\}$, where
$\mathbf{x}_i$ is the location of the $i$th monomer, is given in units of
the LJ energy scale $\epsilon$ by
$E(\mathbf{X})/\epsilon=\sum_{i>j+1}
v_\mathrm{LJ}(r_{ij})+s_r
\sum_{i}v_\mathrm{bond}(r_{i\,i+1})
+s_\theta
\sum_{k}v_\mathrm{bend}(\theta_k)+
s_\tau\sum_{l} v_\mathrm{tor}(\tau_l).$
The dimensionless Lennard-Jones potential with cut-off is given by 
$v_\mathrm{LJ}(r)=4[(\sigma/r)^{12}-(\sigma/r)^{6}]-v_c$ if $r<r_c=2.5\sigma$,
where $r$ is the distance between two nonbonded monomers and
$v_c\approx -0.0163$ is a constant shift to ensure $v_\mathrm{LJ}(r_c)=0$. 
For
$r> r_c$, we set $v_\mathrm{LJ}(r)=0$. Distances are measured in units of
the length scale $r_0$, given by the location of the LJ potential minimum.
The van der Waals radius of a monomer is chosen to
be $\sigma=2^{-1/6}r_0$. The FENE (finitely extensible nonlinear
elastic) bond potential is employed in
the form~\cite{kremer1} $v_\mathrm{bond}(r)=\log \{1-[(r-r_0)/R]^2\}$.
The bond strength is $s_r=-KR^2/2\epsilon$; the parameters
were set to standard values $K=(98/5)\epsilon r_0^2$ and
$R=(3/7)r_0.$ The pay-off for
bending the chain is $v_\mathrm{bend}(\theta)=1-\cos(\theta-\theta_0)$,
where $\theta_0$ is the bond angle in the ground state. The bending energy
scales with
$s_\theta=S_\theta/\epsilon$. For the simulations of the flexible polymer
$S_\theta=0$ (no bending restraint), whereas for the semiflexible polymer
$S_\theta=200\epsilon$ was chosen. Eventually, the torsion potential is 
$v_\mathrm{tor}(\tau)=1-\cos(\tau-\tau_0)$, with the dihedral torsion
angle $\tau$ and its equilibrium value $\tau_0$. The relative energy scale
is $s_\tau=S_\tau/\epsilon$. 
The choice of reference angles $\tau_0=0.873$ and $\theta_0=1.742$
allows for helical segments in the ground-state structures that resemble
right-handed $\alpha$ helices
with about 4 monomers per turn. 
In the following, $\epsilon$, $r_0$, and $k_\mathrm{B}$ are set to unity.
For the simulation of polymers with up to 60
monomers, replica-exchange Monte Carlo parallel
tempering simulations
were performed~\cite{sw1,huku1,huku2,geyer1}.

The propensities of polymers with ($S_\theta>0$) and
without bending restraint ($S_\theta=0$) to form stable helical structures
are investigated under thermal conditions controlled by
the canonical heat-bath temperature $T$. The variation of the torsion
strength $S_\tau$
enables the study of an entire class of helical polymers. Representations
of transition channels in generalized ensembles have turned
out to be beneficial~\cite{mb1,sbj2,jbj1,jbj2}. Therefore, we discuss the
folding channels of the helical polymers in the
multiplicative canonical ensemble provided by the parallel tempering
method.
We 
introduce a pair of order parameters
which
are effectively defined by the average total energies per monomer of the
nonbonded LJ interactions between all monomers and their neighbors up to
6 bonds away, 
\begin{equation}
q_1(\textbf{X})=\epsilon\frac{1}{N}\sum\limits_{i=1}^{N-2}\sum\limits_{j=i+2}^N 
\Theta_{6,j-i}\,v_\mathrm{LJ}(r_{ij}), 
\end{equation}
and all others, 
\begin{equation}
q_2(\textbf{X})=\epsilon\frac{1}{N}\sum\limits_{i=1}^{N-2}\sum\limits_{j=i+2}^N 
\Theta_{j-i,7}\,v_\mathrm{LJ}(r_{ij}), 
\end{equation}
where $\Theta_{kl}=1$ if $k\ge l$ and zero otherwise.
In a single long helix, $q_1$ is minimal and $q_2$ maximal, whereas 
for helix bundles with increasing number of segments $q_2$ gets smaller
and $q_1$ larger.

\revis{For a 40mer, Fig.~\ref{fig:dist} depicts for
a selection of
torsion strength values
$S_\tau$ the distributions of conformations found in the generalized
ensemble that covers the temperature interval 
$T\in [0.1,2.0] \epsilon/k_\mathrm{B}$ in $(q_1,q_2)$ space. The left
column
of figures [(a)-(d)] shows the helical transition pathways for the
bending-restrained semiflexible polymer ($S_\theta = 200\epsilon$) and the
right column [(e)-(h)] for the unrestrained flexible polymer
($S_\theta=0$).
The lightgray region represents the area in $(q_1,q_2)$ space, in
which conformations were found at all temperatures and torsion
strengths in the simulations. This 
distribution, which is independent of $T$ and $S_\tau$, 
gives a first impression of the} 
\revis{differences of the conformational phases of
entire classes of semiflexible and flexible polymers. It
depicts the possible folding channels for the polymers. From the figure, it
is obvious that the distributions
spread out much more for the bending-restrained polymer. Individual
sections (phases)
are clearly separated, with less-populated regions in-between. 
This is different
in the case of flexible polymers. Although conformational phases can be
identified as well, their separation is much less prominent. These
differences can be interpreted in the way that in the case of
semiflexible helical polymers, structural phases are more stable, because
these are surrounded by entropically suppressed regions, which cause
free-energy barriers and phase separation between the helical phases.
The black regions in the figures represent the populations in $(q_1,q_2)$
for fixed $S_\tau$ values, i.e., for individual polymer systems and
confirm that semiflexible polymers with a certain torsion strength prefer
to form tertiary structures inside a distinct helical phase only, which is not 
the case for
flexible polymers.}
\begin{figure}
\centerline{
\includegraphics[width=8.8cm]{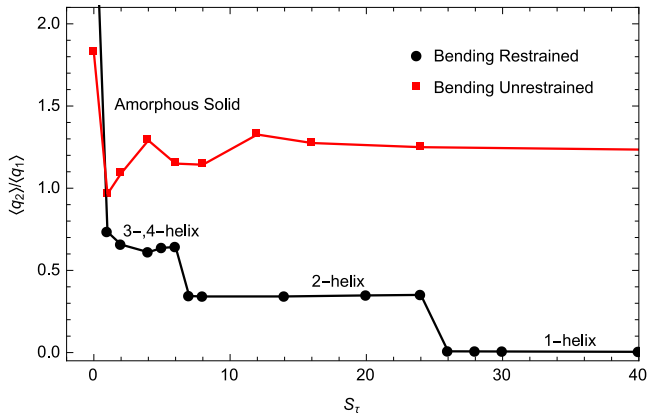}
}
\caption{\label{fig:q1vsq2}%
\revis{Ratio of order parameters $\langle q_2\rangle/\langle q_1\rangle$
for
lowest-energy structures at various
values of $S_\tau$ in the cases of restrained and unrestrained bending.}}
\vspace*{-5mm}
\end{figure}
\begin{figure*}
\centerline{
\includegraphics[width=6.5in]{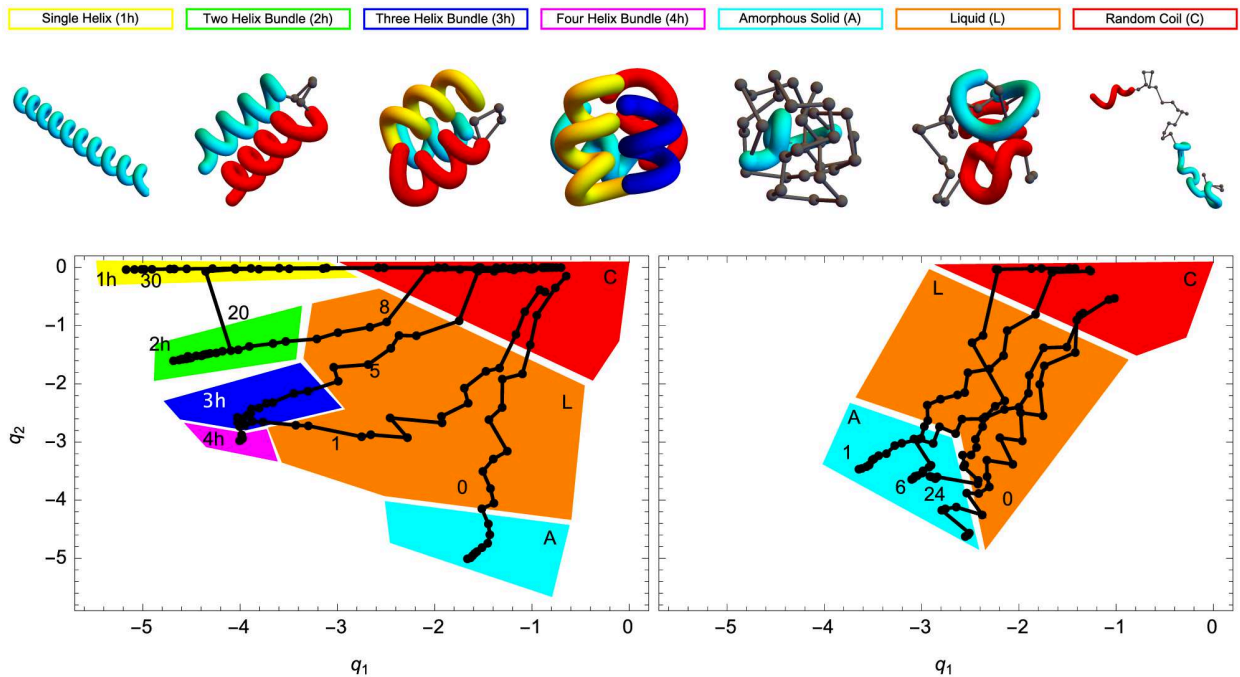}
}
\caption{\label{fig:StructureCharts}%
Structural phase diagrams for bending-restrained semiflexible (left)
and
unrestrained flexible polymers (right) in $(q_1,q_2)$ order parameter space
for the temperature and torsion strength space $(T,S_\tau)$ covered in our
simulations. Colored regions represent structural phases. Black dots locate
free energy minima at given $T$ and $S_\tau$ values. Trajectories
show the
helical folding pathways at fixed torsion strengths $S_\tau$ by decreasing
the temperature.
}
\vspace*{-5mm}
\end{figure*}
%
%
%
%

\revis{The differences in their structural behavior are also clearly
visible when plotting the order parameter ratio $\langle q_2\rangle/\langle
q_1\rangle$ for the lowest-energy structures found in the simulations, as
shown in Fig.~\ref{fig:q1vsq2}. The step-like decrease of this ratio for
the bending-restrained polymers enables the location of the
threshold values of $S_\tau$ where structural phases are separated.}

For a more systematic analysis of the folding behavior and its dependence
on
the torsion strength $S_\tau$ and temperature $T$, 
we define the 
free-energy 
in order parameter space by
$F_{S_\tau,T}(q_1,q_2)=-k_\mathrm{B}T\log Z_{S_\tau,T}(q_1,q_2),$
where 
$Z_{S_\tau,T}(q'_1,q'_2)=\int
{\cal D}X\delta(q'_1-q_1(\mathbf{X}))\delta(q'_2-q_2(\mathbf{X}))
\exp[-E(\mathbf{X})/k_\mathrm{B}T]$
is the restricted partition function in the space of all polymer
conformations 
$\{\mathbf{X}\}$.
Fixing $S_\tau$ and $T$, the free energy
$F_{S_\tau,T}(q_1,q_2)$ 
possesses a global minimum at order parameter values 
$(q_1^\mathrm{min},q_2^\mathrm{min})$. The ensemble of all conformations with 
these order parameter values represents a dominant macrostate of the
system.
The space of macrostates that share a characteristic structural feature
such 
as the number of helical segments in a helix bundle forms a structural
phase.
\begin{figure}[b]
\centerline{
\includegraphics[width=3.6in]{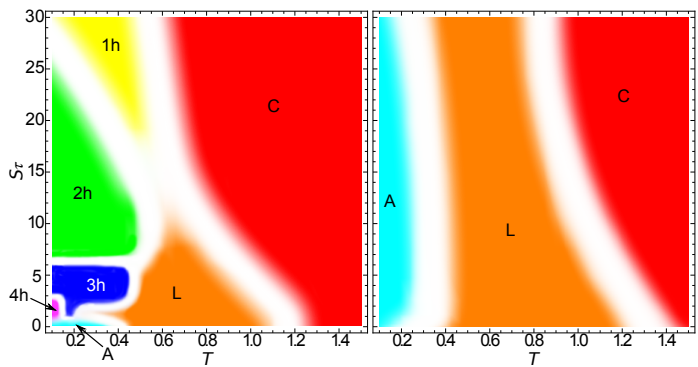}
}
\caption{\label{fig:HyperPhaseDiagram}%
Hyper-phase diagrams of
bending-restrained semiflexible (left) and
unrestrained flexible polymers (right) with $40$ monomers, represented in
the space of the 
torsion strength $S_\tau$ as a material parameter distinguishing classes
of 
polymers and the temperature $T$ as an external control parameter for the 
formation of structural phases. The color code is the same as in
Fig.~\ref{fig:StructureCharts}.}
\vspace*{-5mm}
\end{figure}

\revis{Transitions
temperatures for the various model parameter settings were identified by
standard canonical analyses of extremal fluctuations of energy (specific
heat), structural quantities (e.g., radius of gyration), and order
parameters ($\langle q_1\rangle$, $\langle q_2\rangle$, number of
helices). This will be discussed in more detail elsewhere~\cite{wb1}. Based
on this
information, the $(q_1,q_2)$ space can be separated into regions
(``structural phases'') as shown in Fig.~\ref{fig:StructureCharts} for
bending-restrained semiflexible (left figure) and unrestrained, flexible 
polymers (right figure).} Black lines represent folding trajectories for
several
single polymers with fixed torsion strengths in the interval $S_\tau\in[0,30]$ 
in 
$(q_1,q_2)$ space upon cooling. All trajectories begin in the high-temperature, 
random-coil phase (upper right corner in Fig.~\ref{fig:StructureCharts}, i.e.,  
large $q_1,q_2$ values) and propagate toward a helical state by decreasing the 
temperature. The folding channels at given $S_\tau$ values effectively
connect 
free-energy minima (dots) at various temperatures. 

The 
structural phases of flexible polymers are less well separated, in which
case folding 
channels, after passing a liquid phase, end in the solid amorphous phase.
This general transition behavior is virtually independent of the torsion
strength 
$S_\tau$.
However, the influence of the value of $S_\tau$ upon helix and helix-bundle 
formation is significant for the bending-restrained, semiflexible polymer. 
For $S_\tau=0$, the behavior is similar to that of the flexible
polymer,
but for torsion strength $S_\tau=1$, it
crystallizes initially into a three-helix bundle and then undergoes a
solid-solid 
transition. It emerges from it as a four-helix bundle, which is
energetically 
slightly more favorable at very low temperatures. 
Three-helix bundles clearly form at sufficiently large torsion strength
(e.g., for $S_\tau=5$).

Increasing the torsion strength further favors the extension of helical 
segments, compared to local nonbonded contacts.
The torsional interaction overcompensates what had been an 
energetic gain of nonbonded monomer-monomer contacts despite necessary bending 
penalties. The number of turns is reduced to a single one and a double-helix 
forms in the solid phase ($S_\tau=8$).
Bending-restrained polymers with 
$S_\tau=20$ coexist in a 
transition state between single- and double helix, i.e., in an intermediate 
ordered solid phase a single helix is formed first, which upon further cooling 
splits into a double-helix at the expense of the formation of a single turn. 
This torsion strength marks the threshold at which distant monomer-monomer 
contacts are still formed. For torsion strengths close to $S_\tau=30$ and 
beyond, only stable single-helix phases form.

The complete
structural hyper-phase diagram, parametrized by temperature $T$ and torsion
strength $S_\tau$, is depicted in 
Fig.~\ref{fig:HyperPhaseDiagram} for bending-restrained, semiflexible
polymers (left) and for unrestrained, flexible polymers with torsion 
(right). The phase
diagram for the semiflexible polymer exhibits apparently more structure in
the folded regime at temperatures $T<0.5 $ over the entire interval of
torsion strengths. Whereas at torsion strengths $S_\tau < 7$ four-helix
bundles, three-helix bundles, and amorphous conformations compete and the
phases sensitively depend on the temperature, two-helix bundles and
single-helix conformations are clearly dominant for $S_\tau > 7$.
Remarkably, the liquid (globular) phase disappears for sufficiently large
torsion strengths ($S_\tau>15$), in which case direct coil-helix
transitions occur. Within the range $15<S_\tau<27$, helix-helix
(solid-solid) transitions are present, where single helices
collapse into two-helix bundles by forming a turn. Once the torsion
strength dominates over nonbonded monomer-monomer interaction, i.e., for
$S_\tau>27$, the well-known direct transition from random coils into single
helices occurs.

Contrarily, the folding
process of flexible polymers is hardly affected qualitatively by torsional
constraints [Fig.~\ref{fig:HyperPhaseDiagram} (right)]. The three phases of
random coils, globular, and amorphous
structures are well separate, but 
a helical phase is nonexistent. Furthermore, 
if bending is \emph{not} restrained,
the liquid phase does not disappear and thus a helix-coil
transition does not occur.

In this Letter, we have systematically investigated the influence of
bending restraints
upon the formation of stable helical phases. 
We determined all structural phases for entire classes of flexible
and semiflexible polymers with torsion. These results were summarized in
structural hyperphase diagrams for both polymer classes. 

\revis{The primary result our study is that an effective bending restraint
along the polymer chain is necessary to stabilize helical structures and,
in particular, helix bundles. Different helical structure types that are
separated by entropic gaps in conformational space can only be identified
clearly for semiflexible polymers, whereas for flexible polymers
torsional barriers alone are not sufficient to stabilize individual helical
phases.}

The outcome of this study provides evidence for the natural preference
and significance of locally ordered helical secondary structures for
semiflexible biopolymers, which effectively include DNA and most proteins.
Our results support the understanding of the almost strict confinement of
bond angles in polypeptides (such as bioproteins), which reduces the set of
degrees of freedom that participate in their functional structure
formation to dihedral angles. 
For this reason, it is unlikely that
flexible polymers, i.e., polymers without bending restraint, can be vital
and functional in a biological system. 

This work has been supported partially by the NSF under Grant No.\
DMR-1207437 and by CNPq (National Council for Scientific and Technological
Development, Brazil) under Grant No.\ 402091/2012-4.
\end{document}